\documentclass[12pt, english, twoside]{article} 

\usepackage[letterpaper]{geometry}
\usepackage[titletoc,title]{appendix}

\usepackage{fancyhdr} 
\fancypagestyle{plain} 
{
\fancyhead[LCR]{} 
\fancyhead[C]{}
     
}

\renewenvironment{abstract}
 {\small
  \begin{center}
\normalsize  \textnormal{ABSTRACT\\} \vspace{-0em}\vspace{0pt}
  \end{center}
  \list{}{%
    \setlength{\leftmargin}{0in}%
    \setlength{\rightmargin}{\leftmargin}%
  }%
  \item\relax}
 {\endlist}

\usepackage{sectsty}
\allsectionsfont{\large\raggedright\centering}

\usepackage[]{tocloft}
\addtocontents{toc}{\cftpagenumbersoff{section}} 
\addtocontents{toc}{\cftpagenumbersoff{subsection}} 

\makeatletter
\renewcommand{\@cftmaketoctitle}{}

\renewcommand{\@makefntext}[1]{%
  \setlength{\parindent}{0pt}%
  \begin{list}{}{\setlength{\labelwidth}{6mm}
    \setlength{\leftmargin}{\labelwidth}%
    \setlength{\labelsep}{5pt}
     \setlength{\itemsep}{0pt}%
      \setlength{\parsep}{0pt}%
      \setlength{\topsep}{-3pt}
    \footnotesize}%
  \item[\@textsuperscript{\@thefnmark}\hfil ]#1
  \end{list}%
}

\makeatother

\usepackage{authblk}

\usepackage{graphicx, listings, setspace, hyperref, verbatim}
\usepackage{amsmath, amssymb, mathtools, esint}
\usepackage{tikz, tabu, tabularx, makecell, gensymb, adjustbox, subcaption, float}
\usetikzlibrary{shapes, shadows, arrows}
\usepackage{textcomp}
\usepackage{mathptmx}

\usepackage[T1]{fontenc}
\usepackage[utf8]{inputenc}
\usepackage{mathptmx}

\newcommand\bs{\begin{singlespace}} 			
\newcommand\es{\end{singlespace}} 		
\newcommand\bq{\begin{quote}\begin{singlespace}\small}	
\newcommand\eq{\end{singlespace}\end{quote}}
\newcommand\be{\begin{equation}} 			
\newcommand\ee{\end{equation}}

\newcommand{\ket}[1]{\left| #1 \right>} 
\newcommand{\bra}[1]{\left< #1 \right|} 
\let\baraccent=\= 
\renewcommand{\=}[1]{\stackrel{#1}{=}} 

\hypersetup{
	pdftitle={Looking for Work in Quantum Thermodynamics},
	pdfauthor={Eugene Y. S. Chua},
	pdfsubject={},
	pdfkeywords={},
	pdfcreator={XeLaTeX},
	pdfproducer={XeLaTeX},
	pdftoolbar=false,	
	pdfmenubar=true,	
	pdffitwindow=false,	
	pdfstartview={FitH},	
	unicode=true,		
	pdfnewwindow=true,	
	colorlinks=true,	
	linkcolor=black,	
	citecolor=black,	
	filecolor=black,	
	urlcolor=blue		
}

\usepackage{filecontents}

\usepackage[main=british]{babel}
\usepackage[babel=true]{csquotes} 

\usepackage[backend=biber, style=ext-authoryear-comp,  maxcitenames=2, dashed=false, url=false, doi=false, eprint=true, giveninits=true, innamebeforetitle=true, maxbibnames=6]{biblatex} 
\DeclareFieldFormat{labeldate}{\mkbibbrackets{#1}} 

\DeclareFieldFormat{postnote}{\mkpageprefix[pagination][\mkcomprange]{#1}} 

\AtEveryBibitem{%
 \clearfield{note}%
  \clearfield{series}%
 \clearfield{number}%
 \clearfield{pagetotal}%
 \clearfield{chapter}%
  }

\renewcommand{\intitlepunct}{\addspace\nopunct} 

\DeclareDelimFormat[bib,biblist]{nametitledelim}{\addcolon\space} 

\DeclareFieldFormat{biblabeldate}{\mkbibbrackets{#1}} 

\DeclareFieldFormat{editortype}{\mkbibparens{\textit{#1}}} 
\DeclareDelimFormat{editortypedelim}{\addspace}
\DeclareDelimFormat[bib,biblist]{innametitledelim}{\addcomma\space} 

\DeclareFieldFormat{eprint}{\printtext{available at}\addspace \textless#1\textgreater} 
  
\DeclareFieldFormat[article,periodical]{volume}{\textbf{#1}} 

\DeclareFieldFormat[inbook, incollection, book]{volume}{\bibcpstring{volume}~#1} 

\DeclareFieldFormat{pages}{\mkpageprefix[pagination][\mkcomprange]{#1}} 

\renewbibmacro{in:}{%
\ifboolexpr{%
test {\ifentrytype{article}}%
or
test {\ifentrytype{inproceedings}}%
}
{}
{\printtext{\bibstring{in}\intitlepunct}}
}

\newcommand{\x}[1]{{\color{black}#1}}
\newcommand{\y}[1]{{\color{black}#1}}

\title{{\huge Looking for Work in \\ Quantum Thermodynamics} \vspace{-1.5ex}}
\author{{\LARGE Eugene Y. S. Chua}\\ 
\normalsize Accepted at \textit{The British Journal for the Philosophy of Science}. \\ Preprint of 10 January 2026. Please cite published version when available.}

\date{\vspace{-5.5ex}}

\begin{document}
\maketitle

\noindent\rule{\textwidth}{1pt}

\vspace{5mm}

\begin{abstract}
  \noindent This paper diagnoses a much-discussed problem in quantum thermodynamics, that of generalizing classical work into the quantum domain. I begin with the no-go theorem of Perarnau-Llobet et al (2017): no universal measurement scheme for quantum work satisfies two intuitive, classically consilient desiderata. I assess this incompatibility as stemming from the measurement problem. Decoherence restores compatibility for all practical purposes, but raises questions about what `universality' should mean and whether any measurement scheme can be `universal'. I consider a different standard of universality -- in terms of ontology -- by defining a trajectory-based notion of quantum work using the quantum potential. While this preserves the classical role of work as the integral of forces over distances, and evades the tension of the no-go theorem, consilience fails elsewhere; no single quantum work concept seems capable of preserving all classical features, raising questions for what it takes for successful generalization of the work concept to quantum thermodynamics.
\end{abstract}

\tableofcontents

\vspace{5mm}
\noindent\rule{\textwidth}{1pt}

\newpage 


\section{Introduction}

Classical thermodynamics treats its core concepts -- heat, work, temperature, pressure, and so on -- as unique physical quantities with rich operational and theoretical meanings.\footnote{By classical thermodynamics, I mean equilibrium thermodynamics, including the Minus-First, Zeroth, First, Second, and Third laws. I take this to be conceptually distinct from, though intricately connected to, statistical mechanics.} For instance, in thermodynamic equilibrium, we may interchangeably define the classical work concept in terms of pressure-volume integrals, force‑distance products, or calorimetric measurements. This conceptual richness underwrites the power of thermodynamic concepts and thermodynamics itself. Steam engines, chemical batteries, and compressed springs all do work and give off heat, albeit via different processes. In such contexts, the properties picked out by these concepts are as real as they get.

It may therefore be surprising that the situation is rather different in quantum thermodynamics (QTD). Here, one finds claims that the classical concepts of equilibration \parencite{Yunger_Halpern_2020}, temperature \parencite[\S3.5]{Puglisi_2017}, and entropy production \parencite{Upadhyaya_2024} break down, consilient as they were in classical thermodynamics. Along these lines, \textcite[3]{Mehboudi_2019} note that QTD is concerned with ``... the challenge of consistently redefining the usual thermodynamic variables (i.e. ‘heat’, ‘work’, or ‘temperature’), or even reformulating the laws of thermodynamics to make them applicable to systems which are not macroscopic, but fully quantum." In particular, QTD practitioners report an apparent urgency to the challenge of finding a quantum analogue of the classical concept of \textit{work}. \textcite{Jarzynski2015work} notes ``the absence of a broadly agreed-upon `textbook' definition of quantum work", while \textcite{TalknerHanggi2016aspect} notes that ``there is an urgent need of a theoretical foundation of what work means in quantum mechanics and how it can be defined in an operational way." Similarly, \textcite{lostaglio2018} identifies this lack of agreement on a definition of work as ``one of the main challenges" in generalizing classical thermodynamic and statistical mechanical results to distinctively quantum (i.e. coherent) processes. 

This conceptual challenge was recently rendered precise in \textcite{Perarnau2017nogo}'s widely discussed no-go theorem (\textbf{NGT}). The \textbf{NGT} argues against the very idea of a `universal' concept of (expected) work, where this search for universality is implemented as a search for a ‘universal measurement scheme’. As \textcite[1]{PintoSandGK2024} summarize: ``Remarkably, there is still no statistical framework for describing work that aligns with [the \textbf{NGT}'s conditions], suggesting an incompatibility between quantum mechanics, its convergence to the classical limit, and thermodynamics." 

This paper is a diagnostic of this conceptual situation. I'll argue that the \textbf{NGT} highlights a conflict between two classical features of work that cannot jointly obtain in quantum regimes, a conflict best seen as a process-level measurement problem. From this perspective, I suggest that we should not expect the two conditions of the NGT to jointly hold anyway -- they are physically distinct conditions. Decoherence narrows the gap between the two conditions, but only for all practical purposes. I then develop an account of quantum work in terms of quantum forces over trajectories, contrasting it against \textcite{sampaioetalbohm2018}'s similar proposal. While the account helps to restore agreement between the two features problematized by the \textbf{NGT}, I show that the consilience of classical work nonetheless breaks down: the quantum work thus generalized still fails to retain all the theoretical roles of its classical counterpart. I end by questioning what it would take to successfully generalize work to the quantum domain. 

\section{The work of the classical work}

Since the goal is to generalize the `usual' classical work concept to the quantum domain, let's first consider its various `usual' theoretical roles.

\medskip 

\noindent \textbf{Operationalism.} A founding role of work is \textit{operational}, to facilitate measurements and comparisons of the mechanical effects of various machines \parencite{Cardwell_1967, De_Berg1997}. \y{Historically, the concept of work is what's read off from standardized mechanical tasks that allow us to compare the performance of different engines and devices. One prominent example is Smeaton's comparison of the mechanical effects of undershot vs. overshot water wheels in lifting a weight (cf. \textcite{Morris2018SmeatonVisViva}). To ascribe a certain amount of work to a process is, in this operational sense, to say how that process would register on such standardized mechanical tasks (in \S3, we see contemporary searches for similar `measurement schemes').} 

\medskip 

\noindent \textbf{Forces.} To quantify how much work is being done -- and for work to successfully play its operational role early on -- work was given a \textit{mechanical} role by being defined in terms of forces acting on a system over distances along a process -- equivalently, power delivered to the system over the duration of the process \parencite{galilei_onmotion_1960, parent_perfection_1704, smeaton_experimental_1759}. The work done on a system is the line integral of the net force acting on it along some trajectory $\gamma$ : $\int_{\gamma} \mathbf{F}(\gamma) \cdot d \mathbf{x}$. Note, for now, that this is generally a \textit{path-dependent} definition of the work for general time-dependent forces: without a trajectory $\gamma$ along which determinate forces act, we cannot specify the work done.

\medskip 

\noindent \textbf{Energetics.} Work also plays an \textit{energetic} role by being defined in terms of energy differences in a system along a process \parencite{carnot1783, coriolis1829, sexl_observations_1981, duit_understanding_1981, warren_nature_1982, COELHO20092648}.  \textcite{warren_nature_1982} defines it explicitly as the ``capacity for doing work"; this relationship is enshrined in the so-called \textit{work-energy theorem}. It identifies net work done on a system over some process with the difference between a system's initial and final kinetic energies, $W_{net} = E_K^f - E^i_K$. Having done some net work (or, equivalently in classical physics, having had some net work done on it), the system's kinetic energy is accordingly changed. 
    
In addition to this classical work-energy theorem, we can also find a generalized work-energy theorem in special cases which allows us to connect work to changes in \textit{total energy}. If a system, comprising two subsystems $A$ and $B$, can be assumed to have a \textit{conserved energy}, then the \textit{generalized work-energy theorem} states simply that the external work done \textit{to} $A$, $W^A_{ext}$, in some process by $B$ (equivalently here, the external work done \textit{by} $B$) is the net difference in $A$'s \textit{total} energy before and after the process; equivalently, because of energy conservation, it's the net difference in $B$'s total energy before and after the process.\footnote{See \textcite{geabanacloche_2023} for a straightforward derivation.} That is: $W^A_{ext} = E_A^f - E_A^i = E_B^i - E_B^f = -W^B_{ext}$. In the simplest cases, where the internal forces are conservative,\footnote{Otherwise, the external work is still total change in a system's energy but must include dissipation.} the difference in $A$'s total energy is simply the total changes in kinetic energy as well as potential energy $V$. 

Then, we can relate the generalized work-energy theorem neatly to the original work-energy theorem. External work on $A$ -- and total energy change -- can be cleanly interpreted as the total amount of work -- i.e. total energy -- needed to overcome some amount of potential energy, in order to effect some change in kinetic energy, i.e. in order to effect some net work in $A$: $W^A_{ext} = W^A_{net} + \Delta V_A$.

\medskip 

\noindent \textbf{Equilibrium Thermodynamics.} Work also plays a \textit{thermodynamic} role, by being inter-convertible with heat via appropriate quasi-static processes in regimes of (approximate) thermodynamic equilibrium \parencite{joule_mechanical_1850, mayer_krafte_1842, clausius_moving_1850, thomson_dynamical_1852}. This role of work is enshrined in the First Law of Thermodynamics $dU = \delta Q + \delta W$, or, the change in internal energy of a system is the net difference between changes in the system's heat content and changes in the work done on the system by its environment. This establishes conceptual identity between heat and work, in the sense that both are process-dependent forms of changes in internal energy.

\medskip 

\noindent \textbf{Statistical Mechanics.} Finally, work plays a rich \textit{statistical-mechanical} role, by serving as a bridge between possibly out-of-equilibrium microphysics, in terms of work as energy differences and forces over distances in arbitrary processes, and equilibrium thermodynamics, in terms of work done in quasi-static processes. Following \textcite{gibbs1902elementary}, a common starting point is that (i) a system is in equilibrium with a large external environment (such as a heat bath) with temperature $T$ so that it can be described as part of a canonical ensemble with a stationary Gibbs state $\rho_G = Z^{-1}e^{-\beta H}$,\footnote{$\beta$ is inverse temperature $1/k_BT$.}, and (ii) the system's Hamiltonian has a functional dependence on `external' parameters $a_i$ in addition to `internal' positions $\vec{q}_i$ and momenta $\vec{p}_i$, so that $H(\vec{p}_i, \ldots, \vec{q}_i, \ldots, a_i)$.
    
If the system's kinetic energies depend only on $\vec{q}_i, \vec{p}_i$ and are not affected by the environment except via interactions, variations of $a_i$ can be naturally represented via how they change the interaction term of the system's Hamiltonian, such as the potential energy $V$, so that $\frac{\partial H}{\partial a_i} = \frac{\partial V}{\partial a_i} \equiv  A_i$. Relating all these back to classical equilibrium thermodynamics, we can introduce the \textit{expected energy} of the canonical ensemble of systems, $\langle H \rangle_{\rho_G}$.\footnote{$\langle H \rangle_{\rho_G} = \sum_i \rho_G^i H^i = -\frac{\partial}{\partial \beta}\log Z$. $H^i$ are possible energies and the partition function is $Z = \sum_i e^{-{\beta}H^i}$.} By differentiating $\langle H \rangle_{\rho_G}$, we find a familiar form:\footnote{See \textcite[\S4]{gibbs1902elementary}.} $d\langle H \rangle_{\rho_G} = TdS + \sum_i\langle A_i\rangle_{\rho_G} \; da_i$, where $S$ is the Gibbs entropy $- \sum_i \rho_G^i \ln\rho_G^i$. \textit{If} we identified $S$ with the usual thermodynamic entropy, and $\sum_i\langle A_i\rangle da_i$ as the thermodynamic work done on the system, the equation is formally equivalent to the First Law.\footnote{However, it should be emphasized that this is still an equation about \textit{expectation values}. A standard answer is that \textit{its conceptual equivalence to the First Law only (approximately) holds in the thermodynamic limit}: $N = \infty$ with constant $N/V$.} If the system undergoes a quasi-static, adiabatic process, the change in its expected energy equals the thermodynamic work done on the system, giving a \textit{statistical} work-energy theorem $\langle H \rangle_{\rho_G}^f - \langle H \rangle_{\rho_G}^i \equiv \delta W $.

Crucially, statistical mechanics also allows a precise characterization of how a system's equilibrium thermodynamic characteristics constrain the system's possible out-of-equilibrium fluctuations. This has led to so-called \textit{fluctuation theorems}. One of the most well-known results in this genre is perhaps Jarzynski's equality \parencite{jarzynski1997} (see also \textcite{crooks1999}). For a possibly irreversible process taking a system initially in thermal equilibrium to a state possibly out of equilibrium, the equality tells us that the average exponentiation of work done equals the exponentiation of thermodynamic free energy differences between the initial state and a hypothetical state of thermal equilibrium at the same temperature compatible with the final Hamiltonian: $\langle e^{-\beta W}\rangle_{\rho_G} = e^{-\beta \Delta F}$, where $F = U-TS$ and $\beta$ is the inverse temperature of the system when it is initially in thermal equilibrium, e.g. with a heat bath at temperature $T$. Crucially for later, the work $W$ is the time-integral of $\frac{\partial H}{\partial a_i}$ along some particular Hamiltonian trajectory in phase space. This rests on the classical work-energy theorem: ``the work done on an isolated Hamiltonian system is equal to the change in its energy." \parencite[5]{jarzynski1997} That is, the Jarzynski equality gives us an exact characterization of how statistical features of classical \textit{trajectory-level} work done on a system in a (possibly non-equilibrium!) process are constrained directly by thermodynamic properties of that same system in equilibrium. 

In the appropriate regimes, these various perspectives are consilient: they describe a single concept seen through mechanical, energetic, thermodynamic, and statistical-mechanical lenses, so long as we respect each lens’s domain of validity and their limits. As long as we do so, these perspectives agree with each other, pinning down a single concept. In so doing, consilience tracks a robust reference regime, a regime in which we can agree the concept picks out something `as real as it gets', accessible from a variety of perspectives. From the vantage of this reference regime, consilience is useful for tracking conceptual breakdowns in new regimes, especially by tracking which perspectives fail to generalize.\footnote{See \textcite{Chua2023, chuaandcallender2025} for analyses of consilience in the relativistic thermodynamic domain.}

\section{The quantum work: a no-go theorem?}

To what extent is the consilience of classical work preserved in the quantum domain? Recent work by \cite{Perarnau2017nogo} suggests that the operational and energetic roles of the classical work concept fall apart in the quantum domain. 

Their central system of interest is a system with a quantum state $\rho$ and initial Hamiltonian $\hat{H}$. Under an externally-driven process (in which $\hat{H}$ is varied to $\hat{H}'$ via variations of the classical potential $V$), a unitary operator $U$ maps $(\rho, \hat{H})$ to $(U\rho \, U^\dagger,\hat{H}')$. Any quantum process $\Pi$ can generally be understood this way in terms of $\hat{H}, U, \hat{H}'$. However, unlike in the classical case, they note that
\begin{quote}
    Given this process, there are several approaches to obtaining the statistics of work, namely, the set of [measurement] outcomes $\{W\}$ and their probability distribution $P_W$. \ldots unlike in classical mechanics, \textit{in order to build $P_W$ in quantum physics, one has to specify the measurement scheme through which such knowledge is obtained.} (2017, 1)
\end{quote}
Note: their approach (and much of the QTD literature) is shot through -- albeit often implicitly -- with \textit{operationalism}. Here, I take operationalism to be a particular doctrine about physical concepts and their definition: to define a physical concept is to introduce some observable as a physical quantity; in turn, to do so is to provide a measurement \textit{scheme} for that quantity. 

Roughly, a measurement scheme for a quantity is (i) a set of possible outcomes $\{O_i\}$ together with (ii) a rule that assigns probabilities $P(O_i \mid \rho, \Pi)$ to those outcomes for any target system in $\rho$ and undergoing $\Pi$. Then, (iii) it should also provide a way for physically realizing these outcomes by appropriately correlating system states with pointer states of some measurement apparatus. \textcite[2]{Perarnau2017nogo} consider the most general representation of measurement outcomes in quantum mechanics, positive operator-valued measures or POVMs.\footnote{See \textcite{tumulka_povm_2009} or \textcite[\S1.3]{wallace_emergent_2012} for an introduction to POVMs, and \textcite{busch_quantum_1996} for technical discussion.} The idea is that we should at least be able to find a POVM for work if it is to count among the physical quantities of quantum mechanics.\footnote{This is especially given challenges to defining work as a self-adjoint operator prominently raised by \textcite{Talknerlutzhanggi2007}.} They are noticeably silent on (iii); more on this in the next section. 

With operationalism in the backdrop, \textcite{Perarnau2017nogo} focus on two conditions on an adequate measurement scheme for a quantum work concept, which can be seen as generalizations of features of the classical work. The hope is to find ``...a universal scheme to estimate work so that there is no need for adjusting the measurement apparatus to the initial state." \parencite{Perarnau2017nogo} For \textit{any} state, once we specify $\Pi$, a universal measurement scheme must provide us with the same set of possible outcomes $\{W\}$, together with a rule for defining $P_W$ over these outcomes (with which we can define the expected work $\langle W \rangle$).\footnote{A simple example of a universal measurement scheme: the ($z$-oriented, say) Stern-Gerlach measurement scheme tells us that a spin-1/2 system in any arbitrary state $\rho$ will always be measured to be in either $z$-spin-up or $z$-spin-down -- there is no need to change the possible measurement outcomes depending on the state of the system.} 

\subsection{Setting up the no-go theorem}

The search for a universal measurement scheme for work leads \textcite{Perarnau2017nogo} to search for a unique work POVM, $\{M^{(W)}\}$ with $\sum_{(W)} M^{(W)} = \mathbb{I}$, where each $M^{(W)}$ is associated with some measured value $W$ of work.\footnote{In the simplest kind of POVMs, the members of $\{M^{(W)}\}$ are mutually orthogonal projective operators associated with coarse-grained eigenstates of work, resulting in `projection-valued measures' or PVMs. In general the members of $\{M^{(W)}\}$ may have mutual overlaps, and this allows us to discuss imprecise measurements of quantities such as phase-space POVMs (cf. \textcite[Box 1.1]{wallace_emergent_2012}).} Then, the probability, $P_W$, of measuring the system doing some value of work $W$ is $P_W = \text{Tr} (\rho_\psi M^{(W)})$, where $\rho_\psi$ is the quantum state. Now, of interest is whether we can find an \textit{expected work operator} $\mathbb{W}$ which gives us the expected work done for any system, \begin{equation}\label{workoperator}
    \mathbb{W} = \sum_W W M^{(W)}\; ,
\qquad
\langle W \rangle_{\rho_\psi} = \text{Tr} (\rho_\psi\mathbb{W}).
\end{equation} 
Given this set-up, we can motivate some quantum desiderata based on the consilient roles of the ``usual" classical work. In particular, \y{when discussing energy, two roles of classical work are often conflated in virtue of the classical work-energy theorems: an ontological feature associated with the energetic role, and an operational feature associated with the operational role.}

First the ontological feature: classically, if a system is closed, i.e. allowed to exchange energy but not matter with its environment (e.g. via a classical external potential), then the differences in a system's energy -- as it evolves under classical Hamiltonian dynamics -- is equivalent to work done on that system. This is just the work-energy theorem from \S2. A natural quantum generalization of classical Hamiltonian dynamics is the unitary dynamics generated by a Hamiltonian operator $\hat{H}$, and one generalization of the classical state is the quantum state $\rho_\psi$ which will generally have off-diagonal components representing quantum coherence. Then, one way to demand a quantum analogue of the classical work-energy theorem is to substitute differences in energies with differences in expected energies,\footnote{Such a move typically relies on reasoning to do with the correspondence principle.} and equate it to the expected work:
\begin{equation}\label{quantumworkenergy}
    \langle W\rangle_{\rho_\psi} = \langle E_{final}\rangle_{\rho_\psi} - \langle E_{initial}\rangle_{\rho_\psi} =  \text{Tr}(U \rho_\psi U^\dagger \hat{H}') - \text{Tr}(\rho_\psi \hat{H})
\end{equation}
In terms of $\mathbb{W}$, we can write (in the Heisenberg picture):\footnote{\textcite{Perarnau2017nogo}'s formulation employs 
$\mathbb{W} =  \hat{H} - U^\dagger \hat{H}' U$ rather than what I present here, that is, with a difference in sign for the work. To be consistent with my definition in terms of differences between final and initial energies (rather than initial and final energies), I will adopt a different sign convention. Nothing turns on this difference.}
\begin{equation}\label{heisenbergworkenergy}
    \mathbb{W} =  U^\dagger \hat{H}' U - \hat{H}
\end{equation}
This is the \textbf{NGT}'s first condition -- call this the \textit{unitary} condition \parencite[1]{Perarnau2017nogo}.

The classical work-energy theorem also contains a second guarantee. Because measurements can be as non-invasive and gentle as one wants in classical physics (something that warrants caution in quantum mechanics, as we'll see!), one can always learn about the system's intrinsic state as precisely as one wants. This guarantees, in classical physics, that energy measurements of the system at two points in time will give a result arbitrarily close to the initial and final energies. That is, the classical work-energy theorem not only tells us the values of the system's initial and final energies, but also what we would measure \textit{if} the system's initial and final energies were measured. 

This \textit{operational} feature of the classical work-energy theorem is enshrined in QTD by the so-called \textit{two-point measurement scheme} (TPM),\footnote{\textcite{Talknerlutzhanggi2007} attributes this to \textcite{kurchan2001quantumfluctuationtheorem}.} described by \textcite{Talknerlutzhanggi2007} as ``the elementary observation that two energy measurements are required in order to determine the work performed on the system by an external force." To cash this out in quantum mechanics, for some process $\Pi$, we consider projective measurements of the system's initial energy, and then after a period of unitary dynamics due to external variations of the Hamiltonian, a projective measurement of the system's final energy, hence `two-point'. With the initial Hamiltonian, $\hat{H} = \sum_i^N E_i \ket{i}\bra{i}$, the initial projective measurement finds it in some state $\ket{i}$ with energy $E_i$, with Born rule probability $\rho_{ii} = \bra{i}\rho\ket{i}$. Then, the state $\ket{i}$ is allowed to evolve under unitary dynamics $U$. Finally, with the final Hamiltonian, $\hat{H}' = \sum_f^N E_f \ket{f}\bra{f}$, a final projective measurement is made at the end of the process, finding the system in some state $\ket{f}$ with energy $E_f$ with conditional probability $p_{i,f} =|\bra{f}U\ket{i}|^2$. The work $W$ done in this particular process $\Pi$ is now stipulated to satisfy this relation as well, so that $W^{if} = E_f - E_i$. $W^{if}$ occurs with the joint probability that we found the system initially in $\ket{i}$ and then subsequently in $\ket{f}$, that is: $P(W^{if}) = \rho_{ii}\; p_{i,f}$. The TPM is widely used in QTD because it recovers classical stochastic thermodynamical results, such as Jarzynski's equality (cf. \textcite{Jarzynski2015work}) -- more on this later.

As \textcite[3]{Perarnau2017nogo} point out, the TPM for energy differences demands that the probability distribution for particular work values be:
\begin{equation}
    P_{\text{TPM}}(W) = \sum_{i,f} \delta(W - W^{if})P(W^{if})
\end{equation}
That is, the probability of work $W$ is the sum over all probabilities for each $\ket{i}, \ket{f}$ with the same measured energy difference $E_f - E_i$. This can be written down in terms of a POVM with each member satisfying:
\begin{equation}
    M^{(W)}_{\text{TPM}} = \sum_{i,f} \delta (W -(E_f - E_i))\; p_{i, f} \; \ket{i}\bra{i}
\end{equation}
that is, the probability of finding the system to be doing some work in some process $\Pi$, associated with a member of the POVM, is just the sum of probabilities for all the ways that the system can have some possible work value along this process, understood in terms of \textit{measured} energy differences $E_f - E_i$. Then,
\begin{equation}
    \text{Tr} (\rho_\psi M^{(W)}_{\text{TPM}}) = P_{\text{TPM}}(W)
\end{equation}
Thus the operational feature of the classical work-energy theorem inspires a \textit{second condition} -- call this the \textit{TPM condition} -- for \textcite[1]{Perarnau2017nogo}. This is the demand that $\mathbb{W}$ for a quantum system must satisfy the above behavior when the system is put through the TPM: ``for states with no quantum coherence, the results of classical stochastic thermodynamics should be recovered." That is, the probability associated with 
\begin{equation}
    \langle W \rangle_{\rho_\psi} = \sum_W \text{Tr} (\rho_\psi M^{(W)})\, W = \sum_W \text{Tr} (\rho_\psi M^{(W)}_{\text{TPM}}) \, W
\end{equation}
for all $\rho_\psi$ that have decohered in the energy eigenbasis, i.e. have vanishing off-diagonal terms. It's clear that the requirement to understand $M^{(W)}$ in terms of projective initial energy measurements, i.e. a probability distribution over possible energy states $\ket{i}\bra{i}$, and final energy measurements, i.e. similarly for $\ket{f}\bra{f}$, imposes the constraint that off-diagonal terms of $M^{(W)}$ must vanish (cf. \textcite[3, eqs 8-10]{Perarnau2017nogo}).

\subsection{The no-go theorem}

Having established the two conditions, the \textbf{NGT} proceeds by a single counterexample. Consider a two-level system with initial state $\rho_\psi^i$. The process begins with initial Hamiltonian $\hat{H} = \epsilon \ket{1}\bra{1}$, and ends with final Hamiltonian $\hat{H}' = \epsilon' \ket{1}\bra{1}$, with interim unitary evolution given by a specific unitary operator $U = \ket{0}\bra{+} + \ket{1}\bra{-}$, where $\ket{\pm} = (\ket{0} \pm \ket{1})/\sqrt{2}$. Because $\ket{0}$ is the ground state, $\hat{H}\ket{0} = E_0 = E_0' = 0$.

\medskip 

\noindent \textbf{The TPM condition.} Ex hypothesi, there are only two possible energy levels for $i$ and $f$: 0 or 1. Because of the form of $U$, for all $i$, 
\begin{equation}
    |\bra{0}U\ket{i}|^2 = |\bra{1}U\ket{i}|^2 =\frac{1}{2}
\end{equation}
so $p_{i,f} = 1/2$ for all transitions. Then, following the TPM, the possible two-point measured work values from $i \to f$ for $0, 1$, are:
\begin{align}
    & W^{00} = E_0' - E_0 = 0 - 0 = 0 \nonumber, \qquad W^{01} = E_1' - E_0 = \epsilon' - 0 = \epsilon' \nonumber, \\
    & W^{10} = E_0' - E_1 = 0 - \epsilon = -\epsilon, \qquad  W^{11} = E_1' - E_1 = \epsilon' - \epsilon.
\end{align}
Putting these together to get the expected TPM work via \eqref{workoperator}:
\begin{equation}\label{condition2}
    \mathbb{W} = \sum_{i,f} W^{if} \, p_{i,f}\, \ket{i}\bra{i} = \frac{1}{2}\epsilon'\ket{0}\bra{0} + (\frac{1}{2}\epsilon' - \epsilon)\ket{1}\bra{1}\;.
\end{equation}
Notice that $\mathbb{W}$ has vanishing off-diagonal terms here. 

\medskip 

\noindent \textbf{The unitary condition.} Start instead with \eqref{heisenbergworkenergy}: $\mathbb{W} = U^\dagger \hat{H}' U - \hat{H}$. Inspecting $U$, it's clear that:
\begin{equation}
    U^\dagger\ket{1} = \ket{0}\langle +\ket{1} + \ket{1}\langle-\ket{1} = \frac{1}{\sqrt{2}}\bigg(\ket{0} - \ket{1}\bigg) = \ket{-}
\end{equation}
Plugging in our stipulated Hamiltonians and unitary operators, we get:
\begin{align}\label{condition1}
    \mathbb{W} & = U^\dagger \epsilon'\ket{1}\bra{1} U - \epsilon \ket{1}\bra{1} \nonumber = \epsilon'\ket{-}\bra{-} - \epsilon \ket{1}\bra{1}\nonumber \\ &
    = \frac{1}{2}\epsilon'\bigg(\ket{0}\bra{0} - \ket{0}\bra{1} - \ket{1}\bra{0} + \ket{1}\bra{1}\bigg) - \epsilon \ket{1}\bra{1} \nonumber \\ & 
    = \frac{1}{2}\epsilon'\ket{0}\bra{0}  - \frac{1}{2}\epsilon' \ket{0}\bra{1} - \frac{1}{2}\epsilon' \ket{1}\bra{0} + (\frac{1}{2}\epsilon' - \epsilon) \ket{1}\bra{1}
\end{align}
Note that $\mathbb{W}$ has off-diagonal terms in \eqref{condition1}, contrary to $\mathbb{W}$ seen through \eqref{condition2}. It's clear that for any $\epsilon' \neq 0$, the two constraints on $\mathbb{W}$, \eqref{condition1} and \eqref{condition2}, \textit{cannot be true together}. This leads \textcite[5, emphasis mine]{Perarnau2017nogo} to conclude that ``[...] there will probably \textit{never be an equivalently universal notion of a work variable that is independent of the context in quantum mechanics.}" 

It's worth noting that the QTD community has largely accepted the \textbf{NGT} \parencite{mihailescu2023, lostaglio2018, PintoSandGK2024, francica2022}.\footnote{Some workarounds use possibly-negative quasiprobabilities -- sacrificing a straightforward Born-rule interpretation \parencite{lostaglio2018} -- or adopt state-dependent schemes that forfeit the search for a universal measurement scheme \parencite{PintoSandGK2024}. New proposals continue to be developed to this day.} In introducing stochastic versions of thermodynamic quantities such as the entropy and heat, a recent primer on QTD by \textcite[66, emphasis mine]{potts2024quantumthermodynamics} simply asserts, citing the \textbf{NGT}, that ``we do not define a stochastic version of the work... The reason for this is that... there is \textit{no unique way to define stochastic work that obeys all desired properties}."

\section{Working out the measurement problem}

Let us take stock of the conceptual situation. The \textbf{NGT} tells us that we should not assume that the TPM and unitary conditions should both hold for the same physical system. If we did, $\mathbb{W} \neq \mathbb{W}$, an apparent contradiction that is supposed to entail no universal way to define expected work, per \textcite{potts2024quantumthermodynamics}. I propose instead that the \textbf{NGT} simply shows that $\mathbb{W}_{\text{unitary}} \neq \mathbb{W}_{\text{TPM}}$. That is, that the work defined in the TPM setup differs from that of the unitary setup. Putting the situation this way highlights what the alleged contradiction must assume: the setups used to define $\mathbb{W}_{\text{unitary}}$ and $\mathbb{W}_{\text{TPM}}$ must be assumed to be ways of describing the \textit{same} system undergoing the \textit{same} process. That is, the two conditions are assumed to remain \textit{consilient} in the quantum domain. Otherwise, it's not surprising that two different physical processes are described -- by the same concept -- in two different ways. Yet, because these conditions are assumed to jointly hold for the \textit{same process} -- given by $\Pi(\hat{H}, U, \hat{H}')$ -- for the quantum expected work concept, a contradiction is disastrous for defining the concept. 

To my mind, however, the two conditions say rather different things in the quantum domain. \x{The unitary condition tells us to compute changes of mean energy after unitary evolution: notice that in such situations the system can be in arbitrary superpositions in the energy basis, so that there may be no determinate way to ascribe to the system `an energy' at a time, and hence differences in energies over time.} In contrast, the TPM condition is explicitly an operational condition couched in terms of two-point projective measurements, under which systems \textit{always} have a determinate measured difference in energies.


A century of work on the measurement problem drives home this tension between unitary evolution and measurement outcomes.\footnote{Decades of philosophical discussions have expounded the measurement problem, and I will not add much more to it. See e.g. \textcite{Maudlin1995-MAUTMP} for an introduction.} The problem can be simply stated. Given that the system's state is completely described by a quantum state (a wave-function, or a density matrix), the linearity of unitary evolution (e.g. under the Schr\"odinger equation) entails that systems can be in superpositions of states in some eigenbasis, e.g., of $\ket{1}$ and $\ket{0}$ above. However, in apparent contradiction just like the \textbf{NGT}, projective measurements have determinate outcomes: the system is found either in $\ket{1}$ or $\ket{0}$. 

I suggest that the \textbf{NGT} is simply a special instance of the measurement problem -- twice over. The measurement problem is often framed in terms of single-shot measurements (i.e. a single projective measurement). However, since work is done over a \textit{process} occurring over time, two measurements are needed to determine energy differences and how much work has been done (as \textcite{Talknerlutzhanggi2007} emphasized earlier). The issue -- which the \textbf{NGT} highlights -- is now the incompatibility between how quantum mechanics describes a unitarily evolving system -- with possible coherences -- versus how it describes a system on which two projective measurements are made along a process. 

Taking the \textbf{NGT} to be a subspecies of the measurement problem for processes gives us new conceptual resources for addressing it.\footnote{\textcite{franklinseifert2024} does something similar by considering the import of the measurement problem on the problems of molecular structure.} The biggest takeaway is that \textit{all} standard answers agree that the unitary and TPM conditions are generally not physically compatible \textit{no matter what one's favorite approach to the measurement problem is}. To give a simple example, we know that the double-slit experiment is, famously, physically distinct from the double-slit experiment with monitoring. \textit{All} major approaches agree that quantum measurement is an \textit{additional} form of interaction between a large environment and a quantum system which induces decoherence. 

For Bohmian mechanics, the process of measurement changes the system's conditional wave-function by entangling it with the environment, sending the system -- understood as a configuration of Bohmian point-particles -- along different trajectories than if such interaction had not occurred.\footnote{See \textcite{holland_quantum_1993, bohm_hidden_1952_I, bohm_hidden_1952_II, durr_bohmian_1996}.} For Everettian mechanics, the measurement process instead leads to branching of the quantum state, and subsequent emergence of quasi-classical worlds with determinate measurement outcomes.\footnote{See \textcite{wallace_emergent_2012}.} For GRW-type theories, the system-measurement interaction drastically increases the chance that the system's quantum state collapses.\footnote{See \textcite{Ghirardi1986-GHIUDF}.} 

Furthermore, when we consider decoherence,\footnote{See \textcite{sep-qm-decoherence} and references therein.} we see that they are, in fact, \textit{compatible} in a sense, contrary to the \textbf{NGT}.\footnote{As is well-known, decoherence alone cannot solve the measurement problem. But a solution to the measurement problem isn't necessary to show that $\mathbb{W}_{\text{unitary}} \approx \mathbb{W}_{\text{TPM}}$.} This returns us to the earlier silence on condition (iii) for specifying a measurement scheme. If we took TPM measurements seriously as a physical process, the apparatus for projectively measuring work must be something like a pair of quantum systems with large macroscopic degrees of freedom whose energy pointer states $\{\ket{E_i}\}$ are approximately mutually orthogonal (hence suitable for projective measurements). Then, in the counterexample provided in the \textbf{NGT}, any interaction between a system described by $\mathbb{W}_{\text{unitary}}$ and such a large macroscopic apparatus suppresses coherence (i.e. cross-terms) of the $\ket{0}\bra{1}$ and $\ket{1}\bra{0}$ terms on an appropriate (and realistic) decoherence timescale,\footnote{See \textcite{zurek2003decoherencetransitionquantumclassical}.} leading the statistics of $\mathbb{W}_{\text{unitary}}$ to approximate and come arbitrarily close to $\mathbb{W}_{\text{TPM}}$. \textit{If} we were to actually projectively \textit{measure} a system with $\mathbb{W}_{\text{unitary}}$ in the energy basis at two endpoints of the process, i.e. to put it through the TPM procedure, we \textit{would} get statistics approximating $\mathbb{W}_{\text{TPM}}$. That is to say, once we specify what it takes for a system satisfying the unitary condition to be projectively measured under TPM, the unitary and TPM conditions are approximately compatible after all. 

This raises a question for \textcite{Perarnau2017nogo}'s search for a ``universal measurement scheme": \textit{how universal is universal enough}? $\mathbb{W}_{\text{unitary}}$ \textit{is}, for all practical purposes (FAPP) on realistic decoherence timescales,\footnote{For discussion of FAPP, see \textcite{Wallace2010-WALF-3}.} compatible with $\mathbb{W}_{\text{TPM}}$ once the physics of projective measurement is modeled. For systems in any arbitrary state, once we take into account the physics of projective measurements, their statistics are approximately given by $\mathbb{W}_{\text{TPM}}$. Call this \textit{FAPP-universality}.

Granted, FAPP-universality isn't truly state-independence. The \textbf{NGT} still holds per se: the expected work one defines for a unitarily evolving system \textit{alone} -- in virtue of the cross-terms -- still differs from the expected work of systems undergoing TPM-type procedures. However, because the unitary condition did not specify any interaction with any measurement apparatus, it is \textit{underspecified} for the purposes of defining a measurement scheme \textit{in any case}. As \textcite{busch_measurement_2009} points out, any analysis of measurement -- and the sort of measurement outcomes one expects to find -- must include interaction with a measurement apparatus so that $\mathbb{W}_{\text{unitary}}$ \textit{cannot} be the right statistics. 

This highlights a tension for the \textbf{NGT}'s notion of `universality'. Either consider the physics of measurement, or not. If we do, then there is no contradiction contra the \textbf{NGT} -- for the counterexample they provide, $\mathbb{W}_{\text{TPM}}$ consistently describes the expected work value of systems satisfying the unitary condition -- once we include the physics of projective measurement, FAPP -- \textit{and} systems satisfying the TPM condition. If we do not, then the unitary condition alone underspecifies the measurement context and cannot be adequately described by any measurement scheme.  

One might argue that $\mathbb{W}_{\text{TPM}}$'s FAPP-universality isn't really universal, since it is undefined for unitarily evolving systems per se, that is, those not undergoing TPM-type measurements. The question -- something not asked in the literature -- is what sort of work concept defined in terms of measurement schemes \textit{could} be universal beyond FAPP-universality? It would require a measurement scheme that covers situations in which there is no measurement, surely too much to ask of a \textit{measurement} scheme. In \S6 I consider a distinct standard of universality which would cover such situations, by jettisoning the requirement for physical concepts to be defined vis-a-vis measurement schemes. 

\section{Interlude: classical work falls apart in the quantum domain}

The \textbf{NGT}'s most prominent implication -- that the unitary and TPM conditions are incompatible for the purposes of a universal measurement scheme for quantum work -- can be tempered if we settled for FAPP-universality. 

However, what \textit{is} true, and highlighted by the \textbf{NGT}, is that the \textit{consilience} of the classical work concept is \textit{no longer preserved} when we extend work to the quantum domain via $\mathbb{W}_{\text{TPM}}$: in particular, the operational and energetic roles of the work concept fail to agree in full generality.

To make matters worse, \textcite{Hovhannisyan2024energyconservation} observe that the quantum work may further lose the classical work's connection to forces over distances: ``When information about the full statistics of work is required, ... classical mechanics answers by means of trajectories, leaving no prescription for accessing the statistics of work in the quantum regime." If quantum systems do not always have trajectories in the quantum regime, then there aren't always trajectories along which forces are applied, preventing us from defining quantum generalizations of work in terms of forces over distances. As \textcite[6]{sampaioetalbohm2018} point out: ``Quantum mechanics is routinely taught in a way that stresses subjectivity and indeterminism, while abolishing trajectories in phase space." Supposing this `orthodox' view, quantum work simply cannot be defined via the integral of forces over distances, contrary to the classical work.

This raises another problem for the classical work's consilience. The classical statistical-mechanical \textit{work} -- not just \textit{expected} work -- for particular closed systems, which we exponentiate and relate to the free energy in Jarzynski's equality, is given by \textit{trajectory-level work}. However, this rests on (i) conservation of energy along (ii) determinate trajectories. On the `orthodox' view of quantum mechanics, (ii) generally doesn't exist, and we cannot define quantum work via forces over distances. 

The classical consilience of the operational, mechanical, energetic, thermodynamic, and statistical mechanical roles is broken when work is extended to the quantum domain. Is there a way to save this breakdown and restore consilience, without merely appealing to a FAPP-universal notion of work? (I return to FAPP in \S8.) To an extent. As I will argue now, there are approaches to quantum foundations and the measurement problem which provide a definition of quantum work via quantum forces over determinate trajectories. This allows for reconciliation of the ontological and operational roles, while preserving the mechanical role thought to be a lost cause under `orthodox' quantum mechanics. However, as we'll see, this merely shifts the bump in the rug. 

\section{Quantum force over distance: old answers to new problems}

\subsection{Quantum forces at work}

Largely undiscussed by the QTD community (with the notable exception of \textcite{sampaioetalbohm2018}) is an approach to the measurement problem which allows one to define work -- and, derivatively, expected work -- in the quantum regime, contrary to \textcite{Hovhannisyan2024energyconservation}'s claim, by providing tools for defining (generalized) forces-over-distances. Specifically, these theories highlight a feature of non-relativistic quantum mechanics: that it can be interpreted in terms of particle trajectories acted upon by classical and \textit{quantum} potentials. I discuss one such theory, Bohmian mechanics,\footnote{For more on Bohmian mechanics, see \textcite[Ch. 5]{maudlin_quantum_2019} or \textcite[Ch. 7]{Norsen2017-NORFOQ} for an introduction, and \textcite{holland_quantum_1993, durr_bohmian_1996} for classic treatments.} but I note that there is a many-worlds approach -- Newtonian quantum mechanics (also known as the many-interacting worlds (MIW) theory \parencite{halldeckertwiseman2014, Sebens2015-SEBQMA}) -- which also employs the quantum force and entails a formally similar quantum work concept.\footnote{The core \textit{interpretive} difference between the two is that the Bohmian quantum potential encodes the wave-function's action on configurations of Bohmian particles \textit{in a single world}, while the many-interacting-worlds theory's quantum potential encodes interactions between particle configurations in one world \textit{and configurations of particles in other worlds}.} 

Both theories provide a quantum analogue of classical trajectories, and, crucially, the concept of work done along these trajectories whether or not someone measures it -- not just expected work in terms of POVMs and measurement schemes. Because we are not confined to measurement schemes, we have ways of reconciling, more generally, the unitary and TPM conditions that isn't just FAPP. Correspondingly, we have a distinct sense of universality -- a definition of work done on systems in \textit{any} state, not just for TPMs, but in terms of a sharply defined ontology. 

In Bohmian mechanics, contrary to the `orthodox' formalism of quantum mechanics, a quantum system's state is given by the \textit{quantum state} of standard quantum mechanical formalism, represented in the $N-$particle position eigenbasis as a field over configuration space $\mathbb{R}^{3N}$, often represented as a wavefunction $\psi(\textbf{x}_1(t), \ldots, \textbf{x}_N(t), t)$, but also a \textit{determinate configuration} of $N$ particles, each with precisely specified, non-contextual, positions and trajectories given by $\textbf{x}_1(t), \ldots, \textbf{x}_N(t)$.

The unitary evolution of $\psi$ is given by the time-dependent Schr\"odinger equation, 
\begin{equation}\label{schrodinger}
i\hbar\frac{\partial \psi}{\partial t} = \hat{H}\psi = -\frac{\hbar^2}{2} \sum^N_i \frac{\nabla_i^2}{m_i}\psi + V(\textbf{x}_1, \ldots, \textbf{x}_N,t) \; \psi
\end{equation}
with $m_i$ the mass of the $i$\textsuperscript{th} particle and $V$ the classical potential energy. Now, decompose $\psi$ into real-valued functions $R(\textbf{x}_1, \ldots, \textbf{x}_N,t)$ and $S(\textbf{x}_1, \ldots, \textbf{x}_N,t)$, so that $\psi = R\, e^{\,iS/\hbar}$. $R$ is the amplitude of $\psi$ while $S$ is the phase of $\psi$. $S$ plays the role of determining particle velocities via an additional postulate dubbed the \emph{Guidance Equation}: 
\begin{equation}
    \frac{d\mathbf{x}_i}{dt} = \mathbf{v}_i^\psi(\textbf{x}_1, \ldots, \textbf{x}_N, t) \equiv \frac{1}{m_i}\left( \nabla_i S \right),
\end{equation}
That is, the velocity of the $i$\textsuperscript{th} particle is determined by the spatial variations in the phase of $\psi$. This $\psi-$dependence of \textbf{v} is denoted with the superscript $\psi$. 

To connect this theory to the apparent probabilistic observations and measurement theory of `orthodox' quantum mechanics, a further condition of \textit{quantum equilibrium} must be assumed. This requires that the initial configuration \(\mathbf{x}(0)\) is distributed as \(|\psi(0)|^{2}\),
so that equivariance of the continuity equation implies \(\mathbf{x}(t)\sim|\psi(t)|^{2}\) for all \(t\). Hence the Born rule for positions and -- via pointer positions -- the usual measurement statistics.\footnote{This rests on a reductionism about the role of positions in measurement: ``all observations are position observations, if only the positions of instrument pointers" \parencite[166]{bell_speakable_2004}.}

Crucially, in the Bohmian approach, the operational and ontological roles that fell apart in orthodox quantum mechanics are generally reconciled, in the sense that the generalized state of the system -- $\mathbf{x}_1(t), \ldots, \mathbf{x}_N(t)$ and $\psi(t)$ -- determines exactly what will be observed. Then, the density $\rho(t)$ associated with $\mathbf{x}_1(t), \ldots, \mathbf{x}_N(t)$ determines both the probability with which the system \textit{is}, ontologically, in $\mathbf{x}_1(t), \ldots, \mathbf{x}_N(t)$, and operationally, the probability that one observes $\mathbf{x}_1(t), \ldots, \mathbf{x}_N(t)$. What changes in the case of measurement is simply that $\psi$ must include the interaction between a large measurement apparatus and the target system.

For our purposes, the key resource provided by the Bohmian framework is its capacity for defining \textit{work}, observed or unobserved, not \textit{just} observed \textit{expected} work. To my knowledge, only \textcite{sampaioetalbohm2018} has discussed something similar in QTD; I compare our proposals later. 

As is well-known, we can substitute $\psi = R\, e^{\,iS/\hbar}$ into \eqref{schrodinger} to get a pair of coupled partial differential equations of $R$ and $S$. The first equation, for $\partial \rho/\partial t$ can be interpreted as a continuity equation. The second one, for $\partial S/\partial t$, gives us a crucial piece of our definition of quantum work. It is the \textit{quantum Hamilton-Jacobi equation}:\footnote{Bohm later rejected a causal -- mechanical -- understanding of the quantum potential, and nowadays this formulation is often abandoned in favor of the first-order guidance equation. However, see \textcite{guarini_bohm_2003} for a critique of Bohm's reasoning. For the purposes of this paper, I am sympathetic to Holland's complaint against the first-order approach: ``Since the purpose of the causal interpretation is to offer an explanation of quantum phenomena it seems strange to ignore theoretical structures which may aid that objective. The introduction of the quantum potential as a causal agent has explanatory power which one unnecessarily foregoes by concentrating on just [the guidance equation]. It represents the difference between classical and quantum mechanics." \parencite[78]{holland_quantum_1993}}
\begin{equation}
-\frac{\partial S}{\partial t} \equiv  E^\psi = \frac{1}{2}\sum_i^N \frac{(m_i\mathbf{v_i}^{\psi})^2}{m_i} + V + Q^\psi (\textbf{x}_1, \ldots, \textbf{x}_N, t) \label{particleenergy} 
\end{equation} 
where $E^\psi$ is the total particle energy, the sum of each particle's kinetic energy and the total potential energies \parencite[eq. 3.2.20]{holland_quantum_1993}. Importantly, note that this is \textit{not} just the classical Hamiltonian -- the sum of kinetic and classical potential energies. The total particle energy also includes $Q$, the so-called \textit{quantum potential} \parencite[\S7.1.2]{holland_quantum_1993},\footnote{The idea dates back to Bohm's original papers \parencite{bohm_hidden_1952_I}. \textcite{hiley_bohm_2010} emphasizes, ``this term isn't added, it
is already present in the real part of the Schr\"odinger equation".}
\begin{equation}\label{quantumpotential}
    Q^\psi (\textbf{x}_1, \ldots, \textbf{x}_N, t) = -\sum_i^N\frac{\hbar^2}{2m_i}\frac{\nabla_i^2R}{R}
\end{equation}
which, like the Bohmian velocity, depends on $\psi$. In terms of $Q^\psi$, we find:
\begin{equation}
   m_i \frac{d\mathbf{v}^\psi_i}{dt} = -\nabla_i (V + Q^\psi)
\end{equation}
This looks formally analogous to Newton's Second Law: the instantaneous change of a particle's momentum is numerically equivalent to the gradient of the particle's total potential energy at that time, \textit{if} we interpreted $Q^\psi$ as \textit{also} contributing to the particle's potential energy. This leads us to a natural physical intuition: that the particle is interacting with other particles via $V$, and with the \textit{wavefunction} via $Q^\psi$. If we took this seriously as a quantum generalization of Newton's Second Law, then we can equate the quantum \textit{force} acting on the $i$\textsuperscript{th} particle at a time with $-\nabla_i (V + Q^\psi)$: 
\begin{equation}
    \mathbf{F}^\psi_i(\textbf{x}_1, \ldots, \textbf{x}_N) \equiv -\nabla_i (V + Q^\psi) 
\end{equation}
This is the quantum analogue of the classical force $\mathbf{F} = -\nabla V$.

Now, with the quantum analogue of force in hand, and a precise conception of trajectories -- positions over time -- given by $\mathbf{x}(t)$, defining work in this framework is straightforward. We simply generalize the classical expression for work, in terms of force over distance, to include both classical and quantum forces. The \textit{quantum work} $W^M$ ($M$ for mechanical quantum work) done on the $i$-th particle along its trajectory $\gamma$ from times $t_1$ to $t_2$ can be equivalently written down in terms of \textit{quantum power} over time along the particle's Bohmian velocity $\textbf{v}^\psi_i$ (i.e. along its configuration space trajectory): 
\begin{equation}
    W^M_i \equiv  \int_\gamma \mathbf{F}^\psi_i\cdot d\mathbf{x_i} = \int^{t_2}_{t_1}  -\nabla_i (V + Q^\psi) \cdot \mathbf{v}^\psi_i \, dt \equiv \int^{t_2}_{t_1} P_i^\psi \, dt
\end{equation}
where the integrand captures the total work done on the $i$\textsuperscript{th} particle due to \textit{both} classical and quantum forces along the trajectory of that particle. Correspondingly, $P_i^\psi$ is the quantum power (force over velocity) transferred to the $i$\textsuperscript{th} particle and, physically, is the change in that particle's kinetic energy over time. For a system of $N$ particles, the total quantum work is the sum of the work done on each particle:
\begin{equation}
    W^M \equiv \int^{t_2}_{t_1} \sum^N_{i=1} \; P_i^\psi \; dt
\end{equation}
So far we have been working at the level of individual particle trajectories. What if we did not know the individual particle trajectory -- the microstate -- and only had the wavefunction? Given a wavefunction $\psi$ undergoing some process over time, what is the average work we can expect to be done on the particle configuration of the system? 

Start by defining the expected power density $\pi$ of a single particle at a time, which is the power distributed with respect to $\rho \equiv  |\psi|^2$:
\begin{equation}
    \pi_i(\textbf{x}_1, \ldots, \textbf{x}_N, t) \equiv  \rho(\textbf{x}_1, \ldots, \textbf{x}_N, t) \; P_i^\psi \; 
\end{equation}
Then, the expected power at a time on the $i$\textsuperscript{th} particle is simply:
\begin{equation}
    \langle P_i^\psi \rangle_\psi \equiv \int_{\mathbb{R}^{3N}} \pi_i \; d\mathbf{x}_1, \ldots, d\textbf{x}_N
\end{equation}
Then the expected work done on that particle is the total expected power over time:
\begin{equation}
   \langle W_i^M \rangle_\psi = \int_{t_1}^{t_2} \langle P_i^\psi\rangle_\psi \; dt
\end{equation}
and, using the linearity of expectation, the total expected work done on all particles $\langle W \rangle$ is simply the sum of the expected work done on each particle: 
\begin{equation}
    \langle W^M \rangle_\psi = \sum_i^N \langle W_i^M \rangle_\psi
\end{equation}
so that we have both actual trajectory-level work \textit{and} expected work, all from the concept of quantum forces, work, and power, acting on the Bohmian particles in some actual configuration and along some actual trajectory.\footnote{In more realistic settings, suppose the system is prepared as a proper statistical mixture of pure states $\{\psi_j\}$ with classical mixing probabilities $\{p_j\}$ (i.e.\ on each run $\psi_j$ is prepared with probability $p_j$, where $\sum_{j} p_j = 1$ and $0 \le p_j \le 1$). 
In this case, ensemble-averaged quantities are obtained by averaging over the mixture; for example, the expected work becomes $\langle W \rangle_{\{\psi_j\}} \;=\; \sum_j p_j\, \langle W \rangle_{\psi_j}$.
}

I have just provided a definition of quantum work that can be said to be not only FAPP-universal, but \textit{ontologically universal}: it can be defined in every physically conceivable scenario allowed in Bohmian mechanics (mutatis mutandis, for the analogous quantity in MIW).\footnote{As mentioned before, QTD appears to function in the non-relativistic setting so that this limitation isn't a worry for the project I am interested in here. But, if we wanted to generalize quantum thermodynamics to relativity, then the worries for relativistic thermodynamics may return \parencite{chuaandcallender2025, Chua2023}. Furthermore, old worries about the limits of Bohmian mechanics may return too \parencite{Wallace2023-WALTSI-3}. More work remains for the Bohmian framework (see e.g. \textcite{holland_quantum_1993, Struyve_2011} for work in that direction).} $W^M$ can be defined for all quantum systems, measured or unmeasured, coherent or decohered. In this sense $W^M$ avoids the tension that drove the \textbf{NGT}, between coherent states' unitary evolution and the ability to define a measurement scheme for them, by forsaking the requirement that we define our physical concepts only in terms of measurement schemes. Finally, and relatedly, this approach resolves the tension between the operational and ontological roles of work -- it plays both roles, just as in classical mechanics.

\subsection{A tale of two works: energy and (departures from) classicality}

\textcite{sampaioetalbohm2018} also defines a work concept within a broadly Bohmian framework. However, they define it in terms of temporal variations of \x{the sum of the standard Hamiltonian plus the quantum potential}, i.e. of the total particle energy $E^\psi$. It's useful to distinguish their \textit{energy-based} work concept $W^E$, from the \textit{mechanical} $W^M$ I introduced in \S6.1. The energy-based work done on the system, $W^E$, in terms of differences in the total particle energy, is:
\begin{equation}\label{quantumenergywork}
    W^E \equiv \int^{t_2}_{t_1} \frac{\partial E^\psi}{\partial t} dt  \Bigg \vert_{\textbf{x}(t)}
\end{equation}
evaluated, as before, along a Bohmian trajectory in configuration space. In the special case of unitary evolution of a classically closed system,
\begin{equation}
    W^E = E^\psi(\mathbf{x}(t_2), t_2) - E^\psi(\mathbf{x}(t_1), t_1) \label{quantumenergydiff},
\end{equation}
and the work done on that system just is the change in the total energy of a system -- this amounts to a demand that the classical generalized work-energy theorem from \S2 holds: the work done on a system is equal to its change in total energy.

On the face of it, their approach looks rather different, being entirely in terms of the Hamiltonian. However, \eqref{quantumenergydiff} can also be shown from the perspective of $W^M$ -- but only as a special case. When evaluated along a particular trajectory in configuration space, expanding \eqref{quantumenergywork},\footnote{Here we use the identity familiar from Hamiltonian mechanics that $\frac{dH}{dt} = \frac{\partial H}{\partial t}$ when evaluated along a solution to the (quantum) Hamilton-Jacobi equations (such as Bohmian trajectories).} we get
\begin{equation}
    \int^{t_2}_{t_1}\frac{\partial E^\psi}{\partial t} \; dt \; \Bigg \vert_{\textbf{x}(t)} = \int^{t_2}_{t_1} \bigg( \; \sum_i^N \mathbf{v_i}^{\psi}\cdot m_i\frac{d \mathbf{v_i}^{\psi}}{d t} + \frac{d V}{d t} + \frac{d Q^\psi}{d t}  \bigg) \; dt \;= \; \int^{t_2}_{t_1} \bigg(\sum_i^N P_i^\psi + \frac{d V}{d t} + \frac{d Q^\psi}{d t} \bigg) \; dt
\end{equation}
If the classical \textit{and} quantum potentials are \textit{time-independent}, then:
\begin{equation}
    \int^{t_2}_{t_1}\frac{\partial E^\psi}{\partial t} \; dt \;  \Bigg \vert_{\textbf{x}(t)}= \int^{t_2}_{t_1} \sum_i^N P_i^\psi \; dt
\end{equation}
and the total work done over time (along a trajectory) using $W^M$ is:
\begin{equation}
   W^M = \int^{t_2}_{t_1} \sum_i^N P_i^\psi \; dt \; =  E^\psi(t_2) - E^\psi(t_1)  = W^E,
\end{equation}
that is, exactly the work $W^E$ defined by \textcite{sampaioetalbohm2018}. As \textcite[3]{sampaioetalbohm2018} show, it also quickly follows that:
\begin{equation}
    \langle W^M \rangle_\psi = \langle E^\psi(t_2) \rangle_\psi - \langle E^\psi(t_1) \rangle_\psi =  \langle W^E \rangle_\psi.
\end{equation}
Connecting to the standard quantum mechanical formalism and the standard Hamiltonian operator $\hat{H}$, 
\begin{align}
    \langle W^M \rangle_{\psi} = \langle W^E \rangle_{\psi} & = \langle E^\psi(t_2) \rangle_{\psi} - \langle E^\psi(t_1) \rangle_{\psi}  = \langle \hat{H}(t_2)\rangle_{\psi} - \langle \hat{H}(t_1) \rangle_{\psi}\nonumber \\ & = \text{Tr}[\rho(t_2) \; \hat{H}(t_2)] - \text{Tr}[\rho(t_1) \; \hat{H}(t_1)]
\end{align}
where $\rho = \sum_j p_j \ket{\psi_j}\bra{\psi_j}$. This then connects both $W^E$ and $W^M$ to the standard quantum formalism, in terms of operators and POVMs. 

Nonetheless, it is clear that $W^E$ and $W^M$ generally diverge at the level of actual trajectories (we return to averages at the end). \x{Note, in particular, that the time-independence of $Q^\psi$ is a very strong assumption, since it requires that the quantum state be stationary (e.g. an energy eigenstate, or an appropriate superposition of degenerate eigenstates).} If $V$ or $Q^\psi$ were time-dependent, $W^E = W^M + \Delta V + \Delta Q^\psi$, \x{and $W^E$ is not equivalent to $W^M$, differing exactly by the changes in $V$ and $Q^\psi$ over time, $\Delta V$ and $\Delta Q^\psi$. That is, generally, $W^M$ is a distinct concept of work from $W^E$.} Even though we both work in the Bohmian paradigm, this divergence arises because we have chosen distinct paths to generalizing the classical work to the quantum domain. They have chosen to start with the generalized work-energy theorem, while my approach stems from the classical work-energy theorem (see \S2). My definition of work, $W^M$, tracks only how quantum forces affect trajectories -- and hence how they affect kinetic energy. This definition of work tracks how energy put into the system via $V$ or $Q^\psi$ affects the ontology of the system -- the particles -- solely via the forces they define. 

This might seem like a small difference. After all, in classical physics, as discussed in \S2, both work-energy theorems can be used interchangeably depending on the perspective we are interested in. If we cared about how much $W_{\text{ext}}$ -- total energy $E$ -- must be put into the system to generate some $W_{\text{net}}$ on the system, then we can readily adopt the generalized work-energy theorem, $W_{\text{ext}} = \Delta E = \Delta K + \Delta V$, that is, the change in total energy in terms of kinetic energy $K$ and potential energy $V$, as the appropriate definition of work. But if we wanted to study how the total work done \textit{actually} affects the system internally, $W_{\text{net}} = \Delta K$ tells us that. It is clear how these two perspectives map neatly onto my proposal and \textcite{sampaioetalbohm2018}'s, with the difference of the quantum potential $Q^\psi$: $W^M$ maps onto $W_{\text{net}}$ while $W^E$ maps onto $W_{\text{ext}}$.

However, in the quantum domain the two notions differ much more drastically in physical meaning. The culprit is precisely $Q^\psi$, which couples the particles to the wave-function in a way that has no Newtonian analogue. Because of that coupling, two closely related departures from the classical work concept follow. 

First, $W^E$, but not $W^M$, will fail to track actual changes in the system's trajectory, the sort of change that operationally motivated the development of classical work. Something that \textcite{sampaioetalbohm2018} mention but doesn't emphasize is that Bohmian particles do not obey a classical energy-conservation law in cases where we might expect such a law -- such as when a system appears classically isolated.\footnote{Nonetheless, as Holland reassures us, we will not observe violations of energy conservation ``in the particular instances where one tests for conservation by performing measurements of energy and momentum" \parencite[286]{holland_quantum_1993}, e.g. when we measure a system with a time-independent potential beginning and ending in some stationary energy state.} Even when the external classical potential $V$ is switched off and the classical Hamiltonian is strictly time-independent, $\psi$ can generally evolve into superpositions whose interference structure keeps $Q^\psi$ time-dependent; in such cases, ``the quantum potential can put energy and momentum into a system" even when no energy and momentum is being put into the system externally \parencite[286]{holland_quantum_1993}. The quantum force $-\nabla Q^\psi$ then performs work on the configuration and the kinetic energy varies according to $\Delta K=\int_{t_1}^{t_2}\!\!-\nabla Q^\psi\!\cdot\!\mathbf v\,dt\;\neq 0.$ As \textcite[119]{holland_quantum_1993} emphasizes, ``classically conserved motion is not conserved quantum mechanically for generic solutions of the free Schrödinger equation’’. Consequently a classically closed system -- one that is supposed to have a conserved energy -- can absorb or release mechanical energy without any external agent acting; the physical intuition here is that the wave-function continues to do work on the particles, even when energy is supposed to be conserved. $W^M$ registers this change directly in terms of net work on Bohmian particles via the quantum force depending on $V$ and $Q^\psi$. However, $W^{E}$ will register zero work done, since the change in kinetic energy will be balanced by a change in $Q^\psi$ so that the total energy change is zero. In this sense, if by energy we mean \textcite{sampaioetalbohm2018}'s generalized Hamiltonian, i.e. $E^\psi$, then it is conserved. But it must be emphasized that this generalized energy contains a part ($Q^\psi$) which can change spontaneously, causing changes in kinetic energy of the Bohmian particles, even without any energy put in via external work, simply under unitary evolution. 

This brings us to a second related issue. \textcite{sampaioetalbohm2018}'s understanding of the power, and work, over Bohmian trajectories in terms of the Hamiltonian is ambiguous. If, on the one hand, we have in mind the work and power we put into varying the external potential $V$ to act on the system, then that does \textit{not} result in their definition of work via $W^E$. In the classical case, the amount of energy we put in, in changing $V$, is precisely the amount of energy that gets inserted into the system, because of the conservation of energy. However, in the quantum case, this classical link between ingoing energy and outgoing energy is broken. Because the third part of the generalized Hamiltonian they consider, $Q^\psi$, depends on $\psi$ through $-\sum_i \hbar^2\nabla_i^2R/2m_iR$, which, in turn, \textit{depends on the classical Hamiltonian}, changes in $V$ do ``double work". An amount of energy put into changing $V$ changes the Hamiltonian via $V$ but \textit{also} additionally via $Q^\psi$, so that $W^E$ loses connection with the classical concept of external work -- the amount of work one must put in, to effect some net work on the system -- it was generalized from (cf. \textcite[80-1]{holland_quantum_1993}). 

On the other hand, if we have in mind the work and power that actually gets transferred \textit{to} the particle configuration -- the actual system -- at the end of the process, then $W^E$ does not track that either because it does not track net work; $W^M$ is what tracks net work via the direct changes to particles' trajectories via the quantum force and power. $W^E$ thus sits at a somewhat uncomfortable conceptual space, tracking neither the energy that must be put into a system to effect some work on the system, nor the amount of net work actually done on it to change its total kinetic energy. 

Granted, other related issues with the quantum potential ultimately affect both $W^E$ and $W^M$. These have broader consequences for whether we want to think of either $W^E$ or $W^M$ as genuine mechanical \textit{work}, let alone \textit{thermodynamic} work, in the quantum domain. In Bohmian mechanics there is no back-reaction of  the particle configuration on $\psi$, even as $\psi$ acts \textit{on} them via the quantum force. Newton's third law is generically violated \parencite[79]{holland_quantum_1993}. Via the guidance equation, $\psi$ pilots the particles. However, the theory contains no reciprocal interaction term through which the particle configuration feeds energy or momentum back into $\psi$. The action–reaction symmetry captured by Newton’s third law is therefore missing. Consequently, work done \textit{on} the particles is not equal-and-opposite to work done \textit{by} the particles, and the excess energy comes from nowhere and goes nowhere -- \textit{ex nihilo}, but certainly not \textit{nihil fit}. This seems to imply a lack of \textit{control} over the energy loss, which suggests that the quantum work concepts developed here may also be understood as a form of quantum \textit{heat}. Nevertheless, my proposed $W^M$ makes this violation explicit because it is explicitly formulated in terms of $\psi$ acting on Bohmian particles via the quantum force, without a corresponding partner force acting back on $\psi$. $W^E$ is simply silent on this matter at face value. 

Finally, the broken link between net work done on a system and the energy put in, because of the quantum potential, spoils the hope of finding exact analogues of classical statistical-mechanical non-equilibrium identities such as Jarzynski’s equality in the quantum regime.\footnote{See \textcite{Hovhannisyan2024energyconservation} for a similar analysis within orthodox quantum mechanics.} \textcite[\S IV.C]{sampaioetalbohm2018} trace this directly to the quantum potential and the fact that ``the energy change along individual trajectories is not zero in general". The starting point of \textcite[2690]{jarzynski1997}'s definition of work is precisely that work done on a system can be entirely conceptualized in terms of external variations $\lambda$ of the \textit{classical} Hamiltonian from $t = 0$ to $t = 1$,
\begin{equation}\label{conservationworkenergy}
    W = \int_{0}^{t_s} dt \,\dot{\lambda}\,
     \partial H_{\lambda}/\partial \lambda \,\bigl(\mathbf{z}(t)\bigr),
\end{equation}
so that, for external variations conceptualized in the limit of an infinitely slow quasi-static process \parencite[2691]{jarzynski1997}, the mechanical work done along every phase space trajectory in some ensemble is related to the free energy difference via $W = \int_{0}^{1} d\lambda \, \bigl\langle \partial H_{\lambda}/\partial \lambda \bigr\rangle_{\lambda} = \Delta F$. 

Now, it's clear that the Bohmian quantum work, be it $W^E$ or $W^M$, will violate \eqref{conservationworkenergy}. On the one hand, $W^E$ uses a \textit{non}-classical Hamiltonian. On the other, $W^M$ explicitly shows that the $Q^\psi$ generally results in net work being done on a Bohmian system \textit{even if no external variations of the Hamiltonian occur} -- equivalently, even if no external work is actually being done on the system. Only in cases of a time-independent $\psi$, and hence time-independent $Q^\psi$, will there be no quantum force, but those are hardly the cases of interest when we consider computations and estimates of work done over time. 

As a result, the classical thermodynamic free energy, in general, will \textit{not} track the actual work done on a system, be it $W^M$ or $W^E$. \textcite[\S IV.C]{sampaioetalbohm2018} shows that the average exponentiated work can diverge for $W^E$ in certain cases as well, contrary to the classical average exponentiated work. They further note that the expected quantum work concepts defined above relative to $|\psi|^2$ -- $\langle W^E \rangle_\psi$ or $\langle W^M \rangle_\psi$ -- will not be straightforwardly related to the free energy $\Delta F$ via Jarzynski-type relations, contrary to the classical case: ``in general there is no relation between the average exponentiated work ... and equilibrium quantities, such as $\Delta F$, in the quantum case". This is because we are averaging over an ensemble of trajectories defined relative to $\psi$ or statistical mixtures thereof, and that ``such ensembles are not related to Boltzmann distributions in any direct manner". The resulting distribution is ``thus in general not of Boltzmann form". 

Although Bohmian mechanics (and MIW) strictly reconciles the operational and ontological roles of work unlike $\mathbb{W}_{\text{TPM}}$, it's now clear why the bump in the rug has only been moved, not removed: the original physical meaning associated with the energetic role, enshrined via the work-energy theorems, is no longer preserved when we define work via the quantum potential -- the consilience between the classical and generalized work-energy theorems breaks down. Furthermore, the connections between the energetic, mechanical, statistical-mechanical, and thermodynamic roles of work collapse. Generalizations of the classical work via $W^M$ or $W^E$ result, again, in the breakdown of consilience. 

\newpage 
\section{What does it take for quantum work to work?}

\x{I have argued that three quantum work concepts -- $\mathbb{W}_{\text{TPM}}$, $W^M$, and $W^E$ -- fail to exactly recover the classical roles of work in different ways. What are we to make of this situation? One straightforward way to proceed is to insist that the classical work concept simply does not survive at all in the quantum regime, a form of \textit{eliminativism} about work. But is there any way to save work -- even approximately -- in the quantum domain, as many in QTD want to do? To this end, it is helpful to sketch a spectrum of regimes, from ones in which all concepts agree to ones in which all concepts disagree. 

Let us start with the easiest case: the classical domain. Here, $Q^\psi$ vanishes: we expect the classical Second Law to be true in its entirety without any quantum contribution, with none of the peculiarities of $W^M$ and $W^E$, leaving us with agreement between the classical and generalized work-energy theorems. Without coherence and measurement back-reaction, the worries of the \textbf{NGT} dissolve and $\mathbb{W}_{\text{TPM}}$ just is the classical statistical-mechanical expected work term supporting the usual results in stochastic thermodynamics, e.g., Jarzynski-type equalities. All three concepts agree, recovering the classical consilience of the work concept. This is, of course, unsurprising.

Furthermore, even in the quantum domain, the three concepts can still agree FAPP if certain conditions are met. First, it is generally true (cf. \textcite[\S3.8.2]{holland_quantum_1993}) that the average energy (including $Q^\psi$) is equivalent to the expected Hamiltonian (excluding $Q^\psi$), even if the actual energies differ. At the level of expectation values, then, the difference in expected energies is the same whether or not we included $Q^\psi$. That is, $\mathbb{W}_{\text{TPM}}$ (once we reconcile the unitary and TPM conditions on this work concept, from the \textbf{NGT}, FAPP) and $W^E$ are equivalent at this level, when we take into account the  measurement / decoherence process: both are characterized in terms of the difference in measured energies.  

Second, to actually measure work (and Jarzynski-type equalities), we must set up the system in an initial state of thermal equilibrium -- in which the density matrix is precisely of Maxwell-Boltzmann form (i.e. the Gibbs state): in such a context, \textcite{sampaioetalbohm2018}'s worry for $W^E$ and $W^M$ about Boltzmann form and the breakdown of Jarzynski's equality is largely mitigated, since the appropriate expectation values should also be taken with respect to the Maxwell-Boltzmann distribution, not just $\psi$ at each time-step. Then, recalling \textcite{Talknerlutzhanggi2007}'s `elementary observation" that we need TPM-type schemes to measure the work done on the system by an external force, we would need to make two precise-enough (e.g. projective) measurements at the endpoints of some process. Both would require coupling a system to a large macroscopic environment or apparatus, leading to decoherence. And, by taking the measurement apparatus and decoherence into account, I've argued in \S4 that the operational and energetic roles of $\mathbb{W}_{\text{TPM}}$ can be approximately reconciled. These two points suggest that FAPP, at the level of averages and expectation values, $\mathbb{W}_{\text{TPM}}$ and $W^E$ can be reconciled, and in turn, they can approximately recover all the classical roles of work.

Finally, what about reconciling $W^E$ and $W^M$? I've already discussed one case in which they are equivalent: the case of exactly time-independent quantum and classical potentials. However, I noted earlier that this is a strong assumption, since it would require the system to always be in an energy eigenstate (and nothing particularly interesting happens as a matter of physics). But, we can again appeal to some relevant experimental contexts in which, FAPP, the quantum and classical potentials are close to being time-independent. In addition to contexts in which TPM-type measurements are being made, these are furthermore thermodynamic contexts in which a system again begins in a quantum Gibbs state of thermal equilibrium -- an ensemble of stationary states -- and evolves under a \textit{quasi-static} thermodynamic process: undergoing only slow changes in $V$, and hence slow changes in $Q^\psi$, while remaining in a Gibbs state at each step along the process. In such classical thermodynamic contexts, $W^E$ and $W^M$ approximately agree with each other on the work done. Furthermore, in such a context, \textcite{sampaioetalbohm2018}'s worry for $W^E$ and $W^M$ about Boltzmann form and the breakdown of Jarzynski's equality is largely mitigated, since the appropriate expectation values should also be taken with respect to the Maxwell-Boltzmann distribution, not just $\psi$ at each time-step. So, at the level of expectation values, in regimes where quantum systems undergo quasi-static thermodynamic processes, measured via TPM, all three concepts can still agree with each other, and to the classical concept of work, even if we included the effects of the quantum potential -- that is, beyond the classical domain. 

What if we wanted to study not only quantum systems undergoing quasi-static processes, but also out-of-equilibrium behavior under non-trivial variations of $V$ and $Q^\psi$? In such contexts, the outlook grows dimmer. On the one hand, it remains true that $\mathbb{W}_{\text{TPM}}$ still restores consilience of the operational and energetic roles of work FAPP, and that $W^E$ agrees with $\mathbb{W}_{\text{TPM}}$ at the level of expectation values. Furthermore, as mentioned, the reason for TPM's success and adoption is precisely the fact that the statistics provided by TPM schemes do recover classical stochastic thermodynamical results, including Jarzynski's equality, in which out-of-equilibrium behavior is precisely the target of study. So, we can still preserve consilience between $W^E$ and $\mathbb{W}_{\text{TPM}}$ at the level of expectation values. 

Yet, even at the level of averages, the expected quantum work given by $\langle W^M\rangle_\psi$ will generically \textit{not} behave classically if we cannot ignore variations in the quantum potential. Relying on the TPM setup again, decoherence helps, but only to an extent. Given the initial projective measurement, on a Bohmian approach, the system's conditional wavefunction decoheres to an approximately localized wave-packet on configuration space, such that $R \to 0$ at infinity. Given this condition, \textcite[\S3.8.3]{holland_quantum_1993} proves that $\langle \nabla_i Q^\psi \rangle_\psi = 0$ so that $\frac{d\langle \mathbf{p}_i\rangle}{dt} = \langle -\nabla_i (V + Q^\psi) \rangle_\psi  = \langle -\nabla_i V \rangle_\psi$. That is, the expected quantum force corresponds to the classical force in line with Ehrenfest's theorem. However, this is not yet enough to ensure that $W^M$ behaves classically. What matters for $\langle W^M\rangle_\psi$ is not just the force, but the \emph{power} delivered along the trajectories throughout the process: $\int_{t_1}^{t_2} \langle -\nabla_i (V + Q^\psi)\cdot \mathbf v_i^\psi\rangle_\psi \, dt = \int_{t_1}^{t_2} \langle -\nabla_i V\cdot \mathbf v_i^\psi\rangle_\psi \, dt
- \int_{t_1}^{t_2} \langle \nabla_i Q^\psi\cdot \mathbf v_i^\psi\rangle_\psi \, dt$.
The first term on the right is exactly the classical expected work. However, the second is a distinctively quantum contribution to $\langle W^M\rangle_\psi$ and is defined in terms of the dot product of force and velocity. Even supposing that $\langle \nabla_i Q^\psi \rangle_\psi = 0$, if the actual quantum force at each point of a process is highly correlated with particle velocities,\footnote{These are cases in which $\nabla_i Q^\psi$ contributes much more to particle velocities than the classical $\nabla_i V$, and can be quite common, as in Bohmian models of double-slit experiments \parencite{Philippidis1979}.} then we cannot ignore $\int_{t_1}^{t_2} \langle \nabla_i Q^\psi\cdot \mathbf v_i^\psi\rangle_\psi \, dt$ and $\langle W^M\rangle_\psi$ will generally \textit{not} approximate the classical expected work. $W^M$ fails to line up with $W^E$ and $\mathbb{W}_{\text{TPM}}$ even at the level of expectation values.

So, beyond quasi-static contexts, additional conditions are needed if we want $\langle W^M \rangle_\psi$ to behave approximately classically. These are conditions for which $\int_{t_1}^{t_2} \langle -\nabla_i V\cdot \mathbf v_i^\psi\rangle_\psi \, dt \gg \int_{t_1}^{t_2} \langle \nabla_i Q^\psi\cdot \mathbf v_i^\psi\rangle_\psi \, dt$ so that the quantum contribution is negligible. This occurs in semiclassical regimes for which the de Broglie wavelength of the particle $\lambda$ is small relative to the scale of variation of the classical potential $V$, such that $\lambda \ll L$.\footnote{These are also contexts in which the WKB approximation holds. See \textcite{Allori_2002} and \textcite[\S6]{holland_quantum_1993}.} In such contexts, the quantum contribution to the quantum force is small relative to the classical contribution, so we can neglect $\int_{t_1}^{t_2} \langle \nabla_i Q^\psi\cdot \mathbf v_i^\psi\rangle_\psi \, dt$ and $\langle W^M \rangle_\psi$ approximates classical work. This regime is restricted but nonetheless of interest for QTD: Jarzynski's derivation of classical stochastic thermodynamic results using $\mathbb{W}_{\text{TPM}}$ is precisely done in such a semiclassical regime \parencite{Jarzynski2015work}. 

Thus, even far from thermal equilibrium, there are physically interesting non-classical contexts in which all three concepts line up and recover, FAPP, the consilience of the theoretical roles of classical work: in the semiclassical regime, at the level of expectation values. These concepts provide different perspectives while agreeing with each other. First, they all fulfill an \textit{operational role}: we can compare the mechanical effects of quantum systems by comparing long-run expected measurement outcomes. Second, $W^M$ provides insight into the \textit{mechanical role}: we can define work in terms of \textit{expected forces over distances}. Third, $W^E$ and $\mathbb{W}_{\text{TPM}}$ provides insight into an \textit{energetic role}: we can define it in terms of \textit{expected energy differences} which, FAPP, line up with the operational role. Fourth, they still play a \textit{thermodynamic role}: we can understand it vis-a-vis \textit{thermodynamic quantities}, if the system is in thermal equilibrium and can be ascribed a quantum Gibbs state; and fifth, a \textit{statistical-mechanical role}: even when far from equilibrium, in the semiclassical regime we can see how expected work over trajectories, in terms of $\langle W^M \rangle_\psi$, is approximately classical and can be related to \textit{fluctuation theorems}, such as Jarzynski's equality, and hence to thermodynamic quantities.}

Granted, as I have argued, the quantum work concept will \textit{generally fail to be consilient}, relative to the classical work, when we consider more distinctively quantum regimes. $\mathbb{W}_{\text{TPM}}$ does not describe the work done by a unitarily evolving system \textit{per se}, measurement back-reaction cannot be arbitrarily minimized so that the operational and ontological features of the energetic role cannot be fully reconciled, and full resolution must confront the measurement problem. Then, in approaches to the measurement problem which include a quantum potential, such as Bohmian mechanics or MIW, we find resolution of the earlier problem -- reconciling the energetic and ontological features -- and also a generalization of the mechanical role of the classical work. However, this only leads to breakdown elsewhere, as the classical relationship between mechanical forces and thermodynamic work is broken. What this analysis suggests is a conceptual limit of the classical work concept: beyond specific FAPP contexts such as the semiclassical regime, the classical work concept falls apart, resulting in fragmentation and many possible extensions, though none satisfies \textit{all} theoretical desiderata for the classical work. 

This, then, would explain why, to this day, there is no consensus in QTD with regards to the correct work concept -- there might be wrong generalizations which fail to appropriately satisfy \textit{any} classical desiderata, but there will be no `one true work concept' which can satisfy \textit{all} classical desiderata in general. This returns us to \textcite{Mehboudi_2019}'s earlier description of QTD as the challenge of making thermodynamic variables `applicable' to fully quantum systems. If breakdown of consilience is all but guaranteed, there needs to be greater clarity within the QTD community as to what would count as success for such a task. If we adopted an operationalist standard of success, restricting ourselves to the FAPP context and the expected work, then it seems that we already have an extraordinary generalization of the classical work -- with full consilience -- into a subsector of the quantum domain. However, as I hope to have argued, if success involves a successfully finding a ``unique way to define stochastic work that obeys all desired properties" of the classical work concept as \parencite{potts2024quantumthermodynamics} suggests, then the ``usual" classical work concept simply does not survive in the `fully' quantum domain. This latter standard of success may be `taking thermodynamics too seriously' \parencite{CALLENDER2001539}, and at the same time, not taking the limits -- and the ``fragility'' \parencite[21]{chuaandcallender2025} -- of thermodynamics, such as the measurement context, the need for equilibrium, and its statistical-mechanical underpinnings seriously enough.  

There are at least three options for proceeding. I have already suggested one above: abandon the work concept in all other domains in virtue of its breakdown of consilience -- what might be called an eliminativist stance (cf. \textcite[1317-1318]{Chua2023}). Alternatively, I see two paths forward for the conceptual project of extending the quantum work concept, so long as we accept that no one quantum work concept will fully preserve consilience. The first path embraces this fragmentation, and the second rejects this fragmentation in favor of something more foundational. 

In favor of the first route, one might adopt a stance similar to \textcite{wilson_imitation_2021}'s on the classical force. As \textcite{brading_review_2023} summarizes: ``... `Force’ fragments into several distinct and more precise concepts, each appropriate for its own domain of application. ... mechanics has moved forward not by unifying the various more precise force concepts under a single, general concept, but by explicitly articulating both the boundaries of the domains of application and the means by which we move across those boundaries without disaster...” Likewise, the work concept may fragment in distinct ways to distinct modelling contexts, depending on how we want to use it. For instance, in the context of modeling classically closed, unitarily evolving, systems, we might embrace $W^E$ or $W^M$ as appropriate quantum fragments of the classical work depending on what features of the system we are interested in, rather than $\mathbb{W}_{\text{TPM}}$. As long as we do not conflate these distinct concepts -- respecting their distinct domains of applicability and how they differ -- while ensuring that they reduce to the consilient concept of classical work in the appropriate reference regimes, there is no problem in embracing fragmentation.

Alternatively, in favor of the second route, one might insist that there is a `one true' work concept, perhaps because there is a preferred solution to the quantum measurement problem. Here I must defer to the rich body of literature surrounding the different solutions, and why we might prefer one over another. But that is work for a different context. 

In both cases, I suspect the conceptual project of extending the quantum work concept will follow the trajectory of many other physical concepts, and, indeed, the early history of the classical work concept itself. To quote \textcite[224]{Cardwell_1967}, it will likely follow ``a process of refinement and clarification of concepts, accompanied by their adoption by virtue of their utility and the insights to which they led." In the search for work in quantum thermodynamics, more work remains to be done.

\numberwithin{equation}{section}
\numberwithin{figure}{section}
\numberwithin{table}{section}
\section*{Acknowledgements}
 
I thank Eddy Keming Chen, Yichen Luo, Nelly Ng, Sai Ying Ng, Katie Robertson, Chip Sebens, participants of the Foundations of Thermodynamics Workshop 2025 at Nanyang Technological University, and participants of the John Templeton Foundation Workshop at Chapman University's Institute for Quantum Studies, for their comments and suggestions. I also thank two anonymous referees for their helpful and constructive feedback. This work was supported by a Nanyang Assistant Professorship grant from Nanyang Technological University. 

\begin{flushright}
\emph{
  Eugene Y. S. Chua \\
  School of Humanities\\ 
  Nanyang Technological University\\
  Singapore, Singapore \\
  eugene.chuays@ntu.edu.sg 
}
\end{flushright}

\printbibliography[title ={References}]

\end{document}